\documentstyle[twoside,fleqn,espcrc2,epsf]{article}

\newcommand{\mpsq}{m_{\pi}^2}
\newcommand{\Ninv}{N_{\rm inv}}

\title{Quark Confinement in Multi-Flavor Quantum
Chromodynamics\thanks{Talk presented by K. Kanaya
at the international conference {\it Lattice 92}, 15--19 Sept.\ 1992,
Amsterdam, The Netherlands.}}

\author{Y. Iwasaki\rlap,\address{Institute of Physics, University of Tsukuba,
        Ibaraki 305, Japan}
        K. Kanaya\rlap,$^{\rm a}$
        S. Sakai\rlap,\address{Faculty of Education, Yamagata University,
        Yamagata 990, Japan}
        and
        T. Yoshi{\' e}$^{\rm a}$
       }

\begin{document}

\begin{abstract}
It is investigated how quark confinement depends on
the number of flavors, $N_f$,
in QCD on the lattice with Wilson quarks.
We strengthen and extend the conclusion reported at {\it Lattice 91}:
(1) For $N_f \le 6$ the finite temperature
deconfining transition/crossover line crosses the chiral limit
at finite $\beta$.
We identify the crossing point for $N_f=2$ and 6 on a
$T=4$ lattice, where $T$ is the lattice size in the temporal direction.
We find the phase transition at the crossing point is continuous for
$N_f=2$, while it is of first order for $N_f=6$.
(2) For $N_f \ge 7$, the $T$-independent deconfining transition
observed at $\beta=0$ extends up to $\beta=4.5$ with the critical quark
mass $m_{\rm quark} = O(a^{-1})$, where $a$ is the lattice spacing.

\end{abstract}

\maketitle

\section{Introduction}
It is well known that if the number of flavors $N_f$ exceeds 17,
asymptotic freedom of QCD is lost.
Then a natural question is whether
there is a constraint on the number of flavors for
other fundamental properties of QCD: quark confinement
and spontaneous breakdown of chiral symmetry.
To investigate this problem,
we study the phase structure of
lattice QCD with degenerate $N_f$ Wilson quarks
in the space
of 4 parameters: $N_f$, $\beta$ (gauge coupling), $K$ (hopping
parameter), and $T$ (lattice size in the temporal direction).
For $T=4$, 6, 8, the spatial lattice is chosen to be $8^2 \times 10$,
while for $T=18$, $18^2 \times 24$.
We generate gauge configurations by the hybrid-molecular-dynamics R algorithm
with molecular dynamics time step $\Delta\tau=0.01$, unless otherwise stated.

In this article we report
the results obtained after {\it Lattice 91},
with brief description of
those \cite{ourLat91} reported at {\it Lattice 91}
to make this article self-contained.

\section{$\beta=0$ case}
For $N_f \ge 7$, a strong first order phase transition
occurs when we vary the hopping parameter.
We denote the transition point as $K_d$.
The behavior of physical
quantities implies that quarks are deconfined and chiral
symmetry is recovered for $K > K_d$.
This transition
is remarkably stable under the increase of $T$ up to 18.
Therefore the transition at $K_d$ is a $T$-independent bulk transition.
Thus when $N_f$ exceeds 7, quarks are
not confined and chiral symmetry is not spontaneously broken
for light quarks even in the strong coupling limit
\cite{ourLat91,ourPRL92}.

For $N_f=6$, on the other hand, we conclude that
quarks are confined in the chiral limit $K_c$ ($K_c=0.25$ at $\beta=0$)
from the following facts:
1) The number of iterations for the quark matrix inversion (by CG), $\Ninv$,
increases and exceeds 10,000 at the first stage
of the simulation for $N_f=6$
at $T=4$ with $K=K_c$ .
This is in clear contrast with the case of $N_f \ge 7$,
where $\Ninv$ is $O(100)$.
This difference can be attributed to the existence/absence of
zero eigenvalues of the quark matrix \cite{ourLat91,ourPRL92}.
We will use this difference later as an indicater to discriminate
the two phases.
2) The finite temperature deconfining/crossover
transition line $K_t(\beta)$ on the $T=4$
lattice crosses the chiral limit line $K_c(\beta)$ at finite $\beta$
as discussed in detail below.

\begin{figure}[htb]
\epsfxsize=7.2cm \epsfbox{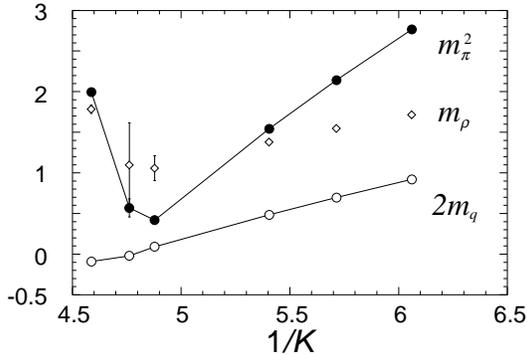}
\caption{Hadron and quark masses at $\beta=4.3$ for $N_f=2$
on the $T=4$ lattice.}
\label{fig:MassFtwo}
\end{figure}
\begin{figure}[htb]
\epsfxsize=6.8cm \epsfbox{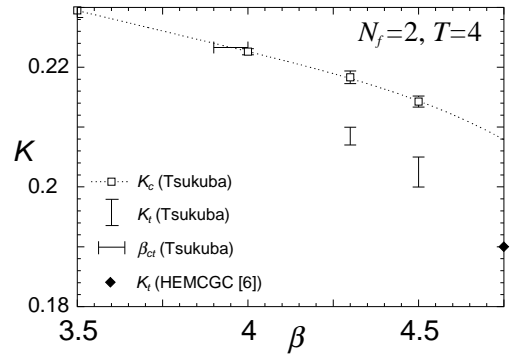}
\caption{Phase diagram for $N_f=2$ at $T=4$.}
\label{fig:PhaseDiagFtwo}
\end{figure}
\begin{figure}[htb]
\epsfxsize=7.5cm \epsfbox{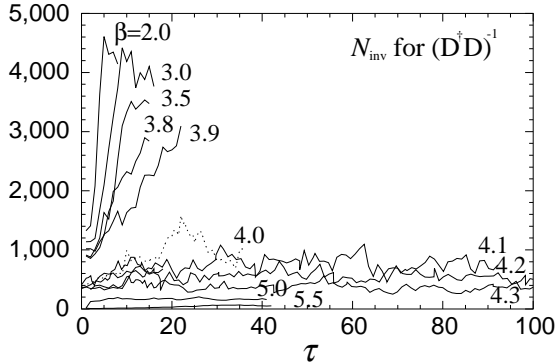}
\caption{Time development of $\Ninv$ for various
$\beta$'s on the $K_c$-line for $N_f=2$ at $T=4$.}
\label{fig:NinvKcFtwo}
\end{figure}

\section{$\beta > 0$ case (i) $N_f \le 6$}
The u and d quarks are very light in the energy scale of QCD
and they are confined.
Therefore we expect that quarks are confined in the chiral limit
for small number of flavors. The results of the
simulations at $\beta=0$ suggest
that this holds for $N_f \le 6$.
Then, the finite temperature deconfining transition line $K_t(\beta)$
should cross the chiral limit line $K_c(\beta)$ at finite $\beta$
for $N_f \le 6$.
However, the confirmation of this is not so easy,
as noted by Fukugita {\it et al.} several
years ago \cite{FOU}.
We have studied this problem both
by following the $K_t$-line to a smaller $\beta$ region and
by monitoring the number of CG iterations, $\Ninv$, on the $K_c$-line.

\subsection{$N_f=2$}
In Fig.\ref{fig:MassFtwo} are shown
the results for hadron masses and $2m_q$ at $\beta=4.3$
for $N_f=2$ on the $T=4$ lattice
by the hybrid MC algorithm.
Here the quark mass $m_q$ is defined through an axial-vector current
Ward identity \cite{bochi,iwasakia}.
We identify the location of $K_t$
by a sudden change of the behavior of physical observables such as $\mpsq$.
$K_c(\beta)$ is defined as the point where
$\mpsq$ in the confining phase vanishes with
a linear extrapolation in terms of $1/K$.
Fig.\ref{fig:PhaseDiagFtwo} is the phase diagram obtained in this way
for $N_f=2$ at $T=4$.

\begin{figure}[htb]
\epsfxsize=6.3cm \epsfbox{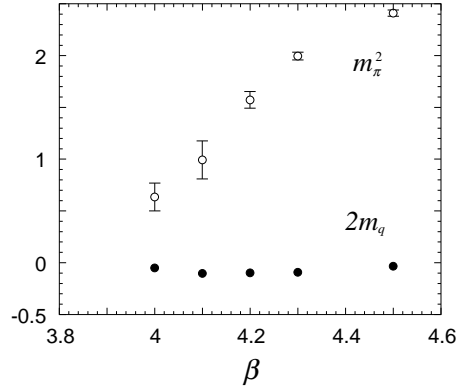}
\caption{$m_{\pi}^2$ and $2m_q$ on the $K_c$-line for $N_f=2$ at $T=4$.}
\label{fig:MassKcFtwo}
\end{figure}
\begin{figure}[htb]
\epsfxsize=6.3cm \epsfbox{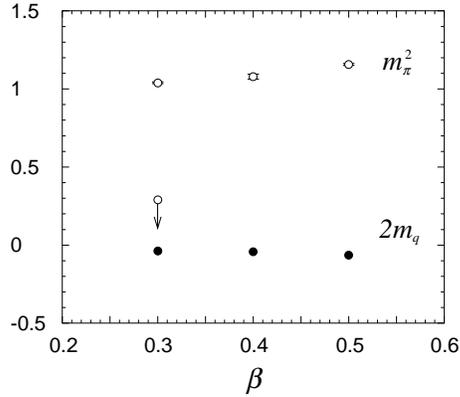}
\caption{The same as Fig.4 for $N_f=6$ at $T=4$.}
\label{fig:MassKcFsix}
\end{figure}
\begin{figure}[htb]
\epsfxsize=6.3cm \epsfbox{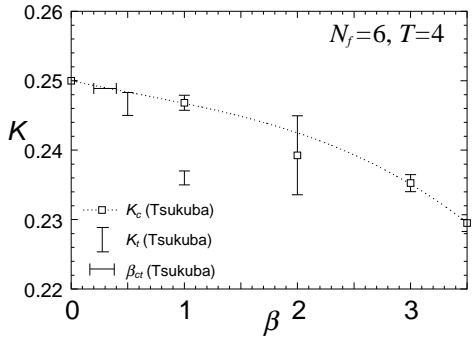}
\caption{The same as Fig.2 for $N_f=6$ at $T=4$.}
\label{fig:PhaseDiagFsix}
\end{figure}

To identify the location of the crossing point, we have done a series
of simulations exactly on the $K_c(\beta)$-line.
Fig.\ref{fig:NinvKcFtwo} is the result of
$\Ninv$ for $N_f=2$ on the $K_c$-line
as a function of the molecular-dynamics time.
We find a clear difference between the cases of $\beta \le 3.9$ and
$\beta \ge 4.0$.
Therefore we identify the crossing point $\beta_{ct}$
as $\beta_{ct}\sim 3.9$ -- 4.0.
This $\beta_{ct}$ is
consistent with the linear extrapolation of the $K_t$-line as
is shown in Fig.\ref{fig:PhaseDiagFtwo}.

To study the nature of the transition at
$\beta_{ct}$, we measure $\mpsq$ and $m_q$ on the $K_c$-line
(Fig.\ref{fig:MassKcFtwo}).
When we decrease $\beta$
toward the crossing point $\beta_{ct}$,
$\mpsq$ decreases linearly
and is consistent with zero at $\beta_{ct}$.
This implies that the chiral phase transition is continuous (second-order
or crossover).

For $T=18$, we find $\beta_{ct} \sim 4.5$ -- 5.0.
Note that the shift of $\beta_{ct}$ with $T$ is small.
This means that if one wants to get a confining chiral limit
for $\beta > 5.0$, one needs $T>18$.

\subsection{$N_f=6$}
Monitoring $N_{inv}$ in the
simulations exactly on the $K_c(\beta)$-line for various $\beta$'s,
we find a two-state signal in the time development
of $\Ninv$ at $\beta=0.3$,
depending on the choice of the initial configuration.
When the initial configuration is chosen to be
a deconfining configuration at $\beta=0.4$,
$\Ninv$ at $\beta=0.3$ is quite stable and small ($\sim 600$)
and $\mpsq$ is large.
On the other hand,
$\Ninv$ for a trajectory starting from a mixed configuration
shows a rapid increase with molecular-dynamics time
and in accord with this, $\mpsq$
decreases with molecular-dynamics time.
Upper bound for $\mpsq$ in this case
is given in Fig.\ref{fig:MassKcFsix},
where the result of $\mpsq$ and $2m_q$ on the $K_c$-line at $T=4$ is shown.
We identify the crossing point $\beta_{ct} \sim 0.2$ -- 0.3,
taking account of other results also.
We see that the results are in clear
contrast with the case of $N_f=6$:
$\mpsq$ is approximately a constant and large
near the crossing point $\beta \sim 0.3$ -- 0.5.
At $\beta=0.3$ we have two values for $\mpsq$
depending on the initial configuration.
The smaller one is consistent with the vanishing of $\mpsq$.
Therefore the chiral transition is of first order for $N_f=6$.
The phase diagram for $N_f=6$ at $T=4$ is shown in
Fig.\ref{fig:PhaseDiagFsix}.

\begin{figure}[htb]
\epsfxsize=7.2cm \epsfbox{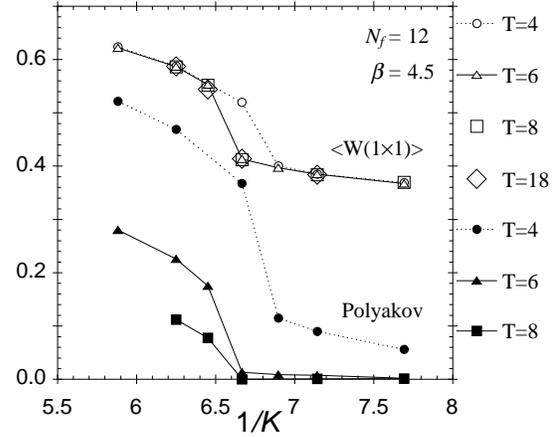}
\caption{Plaquette $W(1\times1)$ and Polyakov loop at $\beta=4.5$
for $N_f=12$.}
\label{fig:PolyPlaqFtwelve}
\end{figure}

\section{$\beta > 0$ case (ii) $N_f \ge 7$}
Repeating similar studies for various $\beta$'s and for various $N_f$'s
to that in the case of $\beta=0$, we
find that the gross feature of the phase transition
is quite similar to that at $\beta=0$.
As a typical example, are shown the results of plaquette and
Polyakov loop 
at $\beta=4.5$ for $N_f=12$ in Fig.\ref{fig:PolyPlaqFtwelve}.
As in the case of $\beta=0$, the deconfining transition point
is stable for changing $T$ for large $T$.
For $\beta \ge 2.0$, however,
we find a small upward $K$ shift of the deconfining transition point
with $T$ for small $T$ ($\sim 4$).

Summarizing these results for $N_f \ge 7$, we have the following phase diagram:
First we have a $T$-independent deconfining transition line $K_d$,
which starts from a finite $K$ at $\beta=0$ and extends to larger $\beta$.
In addition to this line, we have a $T$-dependent deconfining
transition/crossover line $K_t$,
which reaches the $K_d$-line without crossing the $K_c$-line.
With increasing $T$, this $K_t$-line moves toward larger $\beta$.
Up to $\beta=4.5$ for $N_f=12$ and $7$, we find that $K_d$ does not approach
$K_c$ with increasing $\beta$ in the sense that the critical $m_q$
at $K_d$ is O(1) in units of the inverse lattice spacing
and is rather increasing slightly with $\beta$.

\section{Conclusion}
We have studied the phase diagram of QCD with $N_f$ degenerate
Wilson quarks.
For $N_f \le 6$,
the finite temperature deconfining transition/crossover line $K_t$
crosses the chiral limit $K_c$ at finite $\beta$
and shifts toward larger $\beta$ with increasing $T$.
At the crossing point, the data imply that
the transition is continuous for $N_f=2$
at $T=4$,
while it is of first order for $N_f=6$ at $T=4$.
When $N_f$ exceeds 7, we have a $T$-independent deconfining transition
line $K_d$ separating the chiral limit $K_c$ from the confining region.
The finite temperature transition $K_t$-line reaches the $K_d$-line
without crossing the $K_c$-line.
Up to $\beta=4.5$ for $N_f=12$ and $7$, we find
that the critical $m_q$ at $K_d$
is O(1) in units of the inverse lattice spacing.
If this fact holds in the continuum limit, it implies that
quarks are not confined in the continuum limit for $N_f \ge 7$.
Therefore it is important to study the behavior of the $K_d$-line
at larger $\beta$
to conclude the implications in the continuum limit.
A similar bulk deconfining transition is reported recently
in lattice QCD with staggered quarks for $N_f=8$ \cite{columbia}.
It is not clear whether the phenomenon reported is related to
that observed by us.

The simulations
on the $8^2 \times 10 \times T$ ($T=4$--8) lattices
and the $18^2 \times 24 \times T$ ($T$=18) lattice
have been performed with
HITAC S820/80 at KEK and with QCDPAX at the
University of Tsukuba, respectively.
We would like to thank members of KEK
for their hospitality and strong support
and the other members of QCDPAX collaboration for their help.
We also thank Sinya Aoki and Akira Ukawa for valuable
discussions.
This project is in part supported by the Grant-in-Aid
of Ministry of Education, Science and Culture
(No.62060001 and No.02402003).



\begin{thebibliography}{9}
\bibitem{ourLat91} Y. Iwasaki, K. Kanaya, S. Sakai, and T. Yoshi{\' e},
                   Nucl.\ Phys.\ B (Proc.\ Suppl.) 26 (1992) 311.
\bibitem{ourPRL92} Y. Iwasaki, K. Kanaya, S. Sakai, and T. Yoshi{\' e},
                   Phys.\ Rev.\ Lett.\ 69 (1992) 21.
\bibitem{FOU}      M. Fukugita, S. Ohta and A. Ukawa,
                   Phys. Rev. Lett. 57 (1986) 1974;
                   A. Ukawa, Nucl.\ Phys.\ B (Proc.\ Suppl.) 9 (1988) 463.
\bibitem{bochi}    M. Bochicchio, L. Maiani, G. Martinelli, G. Rossi
                   and M. Testa, Nucl. Phys. B262 (1985) 331.
\bibitem{iwasakia} Y. Iwasaki, K. Kanaya, S. Sakai, and T. Yoshi{\' e},
                   Phys.\ Rev.\ Lett.\ 67 (1991) 1494;
                   Y. Iwasaki, Y. Tsuboi and T. Yoshi\'e,
                   Phys.\ Lett.\ 220B (1989) 602.
\bibitem{HEMCGC}   K.M. Bitar, {et al.}, Phys. Rev. D43 (1991) 2396.
\bibitem{columbia} F.R. Brown, {et al.}, Columbia Univ. preprint CU-TP-541;
                   Z. Dong and N. Christ, Nucl. Phys. B (Proc. Suppl.)
                   26 (1992) 314.
\end{thebibliography}
\end{document}